\documentclass{iopart}
\usepackage{epsfig,amssymb}

\begin{document}

\newcommand{\fesn}{Fe$_3$Sn$_2$}
\newcommand{\bftau}{\mbox{\boldmath$\tau$}}
\newcommand{\bfpsi}{\mbox{\boldmath$\psi$}}

\letter%
{Non-collinearity and spin frustration in the itinerant kagome ferromagnet \fesn}

\author{L.A. Fenner$^1$, A.A. Dee$^1$, A.S. Wills$^{1,2,*}$}

\address{$^1$ Chemistry Department, UCL, 20 Gordon Street, London WC1H 0AJ, UK}
\address{$^2$ London Centre for Nanotechnology, 17-19 Gordon Street, London WC1H 0AH, UK}

\ead{a.s.wills@ucl.ac.uk}

\begin{abstract}

Frustrated itinerant ferromagnets, with non-collinear static spin structures, are an exciting class of material as their spin chirality can introduce a Berry phase in the electronic scattering and lead to exotic electronic phenomena such as the anomalous Hall effect (AHE). 

This study presents a reexamination of the magnetic properties of \fesn, a metallic ferromagnet, based on the 2-dimensional kagome bilayer structure. Previously thought of as a conventional ferromagnet, we show using a combination of SQUID measurements, symmetry analysis and powder neutron diffraction, that \fesn\ is a frustrated ferromagnet with a temperature-dependent non-collinear spin structure. The complexity of the magnetic interactions is further evidenced by a  re-entrant spin glass transition ($T_f\simeq80$\,K) at temperatures far below the main ferromagnetic transition ($T\mathrm{_C}$ = 640\,K).

 \fesn\ therefore provides a rare example of a frustrated itinerant ferromagnet. Further,  as well as being of great fundamental interest our studies highlight the potential of \fesn\ for practical application in spintronics technology,  as the AHE arising from the ferromagnetism in this material  is expected to be enhanced by the coupling between the conduction electrons and the non-trivial magnetic structure over an exceptionally wide temperature range.

\end{abstract}

\section{Introduction}

The discovery of unconventional magnetic and electronic phenomena in conductors is important for the development of spintronics:  information technology based on the application and control of electronic spin. The range of mechanisms being enlisted to engineer exotic electronic properties is steadily growing, and includes effects such as the complex quasi-two-dimensional multiband Fermi surface of the Fe-based pnictide superconductors \cite{pnictide1}, centrosymmetry breaking by magnetic order ({\it e.g. } TbMnO$_3$ \cite{TbMnO3}),  double exchange (manganites)\cite{Anderson}, and  competing  interactions between different moment types, {\it e.g.} $d$- and $f$-  moments ({\it e.g.} RECrSb$_3$ series \cite{RECrSb3_1}). One particularly intriguing avenue for research are conductors with non-collinear static spin structures, as their chirality can introduce a Berry phase in the electronic scattering and lead to spin-dependent effects, such as extraordinarily large values of the anomalous Hall effect (AHE). This mechanism for the AHE was first developed by Matl {\it et al.} \cite{MAT98} and Ye {\it et al.} \cite{YE99} to account for the unusual behaviour in La$_{1-x}$Ca$_x$MnO$_3$.  The Berry phase mechanism has also been confirmed to explain the AHE behaviour well in a variety of systems including the spinel CrCu$_2$Se$_4$ \cite{YAO07} and thin films of  Mn$_5$Ge$_3$ \cite{ZEN06}, and has been proposed to account for the AHE that occurs below 100\,K  in the semiconducting pyrochlore Nd$_2$Mo$_2$O$_7$, which features a canted spin-ice-like ferromagnetic spin structure below T$_C$ = 89\,K \cite{TAG01, TAG03}. There is, however, some controversy surrounding the actual mechanism for AHE in Nd$_2$Mo$_2$O$_7$: Yasui {\it et al.} \cite{YAS06} and Sato \cite{SAT07} analysed the magnetic field-dependence of the spin structure, from which they calculated the spin chirality and predicted the Hall resistivity, and found that neither the spin chirality mechanism, nor any of the other currently known AHE mechanisms, can account for the behaviour of this material. This observation reopens fundamental questions over the origin of the AHE in frustrated magnets.

Frustrated magnets, where conventional magnetic order is `frustrated' by a competition between the different magnetic exchange interactions and a large ground state degeneracy, have proven to be one of the simplest domains in which to engineer extraordinary electronic effects. The list of experimentally observed exotic ground states is ever increasing and  includes  the spin glass states of (H$_3$O)Fe$_3$(SO$_4$)$_2$(OH)$_6$ \cite{H3O_anisotropy,H3O_Fe,H30_Fe_2}, SrCr$_{9x}$Ca$_{12-9x}$O$_{19}$ \cite{Limot:2002p65} and Y$_2$Mo$_2$O$_7$\cite{Y2Mo2O7}; the quantum spin liquid states of Herbertsmithite\cite{Herbertsmithite} and Kapellasite (ZnCu$_3$(OH)$_6$Cl$_2$)\cite{Kapellasite}; the spin ice states of bulk Dy$_2$Ti$_2$O$_7$ and Ho$_2$Ti$_2$O$_7$ \cite{spin_ice_1} and the nano-engineered realisations of the spin ices\cite{Artificial_spin_ice,kagom_spin_ice2}. Magnetic frustration can also lead to very rich magnetic phase diagrams, {\it e.g.} for gadolinium gallium garnet (GGG) \cite{GGG}, Gd$_2$Ti$_2$O$_7$ \cite{Gd2Ti2O7_1} and the series Li$_x$Mn$_2$O$_4$ \cite{LixMn2O4_1}. Further, much effort is currently focussed on the degenerate manifold itself as a medium able to support new phenomena, such as order-by-disorder\cite{order_by_disorder}, Kasteleyn transitions \cite{Kasteleyn}, the formation of effective magnetic monopoles \cite{monopoles}, and topological spin glass behaviour \cite{Ritchey:1993p317,H3O_anisotropy,H3O_Fe}. All of these studies are, however, on insulators and progress in the field of frustrated itinerant magnets has been very much hindered by the lack of model systems with which to explore and test the developing theories.

In this article we introduce \fesn\ as a new non-collinear and frustrated itinerant ferromagnet based on a kagome bilayer structure. While the material has been known for many years \cite{Nial} there is much confusion over its magnetic properties, with the analysis of early M\"ossbauer \cite{CAE78, CAE79} and powder neutron diffraction data \cite{MAL78} being hindered by difficulties and inconsistencies. The authors of these early papers concluded that the spins in \fesn\ lie approximately along the {\it c}-axis above 250\,K, and undergo a gradual rotation into the {\it ab} plane below 250\,K, remaining collinear throughout the rotation. Our reexamination of the magnetic properties of \fesn\ followed from the hope that the spins on the Fe-sublattice are actually frustrated, which would lead to characteristic fluctuations and exotic spin-dependent conduction properties. Here, we show using a combination of theoretical and experimental techniques, that spin frustration is both allowed and present in \fesn. Firstly, the presence of spin frustration is indicated by temperature-dependent magnetisation measurements, which reveal the presence of competing magnetic interactions and evidence a re-entrant spin glass component below $\simeq 80$\,K. Symmetry analysis is then applied to demonstrate that ferromagnetism in \fesn\ is not restricted to being collinear, thereby hinting at the rich physics that is possible in this material.  Further, the analysis of powder neutron diffraction data in terms of both collinear and non-collinear magnetic models is presented.

These findings indicate that \fesn\ is a particularly notable candidate for spintronics applications as the high Curie temperature ($T\mathrm{_C}> 600$\,K), and the possible frustration enhancement to the AHE expected for a ferromagnet,  would allow access at room temperature to the effective control of spin polarised currents \cite{SIN08}, as well as providing new routes for the conversion of magnetic data into an electrical signal in devices such as sensors and nonvolatile magnetic memory\cite{GER07}.

\begin{figure}[htb]
\begin{center}
\includegraphics[scale=0.22]{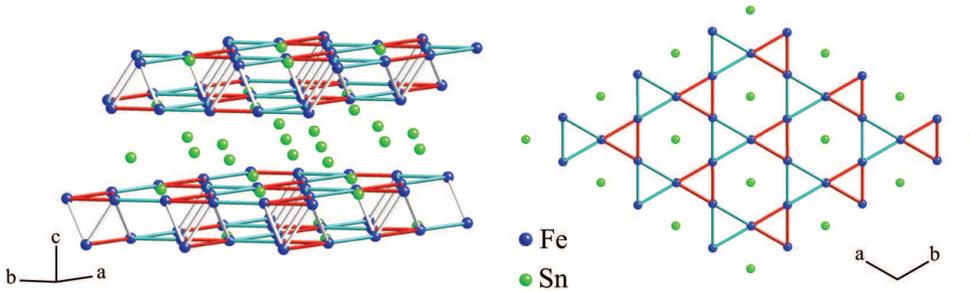}\hspace{0pc}%
\caption{The crystal structure of \fesn\ refined using neutron diffraction data collected on the D20 diffractometer with neutrons of 1.3\AA. The Fe ions form bilayers of offset kagome networks, and the Sn ions lie in the centre of the kagome hexagons and between the bilayers. The blue and red triangles (colour online) indicate the larger and smaller equilateral triangles respectively.}
\end{center}
\label{crystal_structure}
\end{figure}

The crystal structure of \fesn\ is shown in  figure \ref{crystal_structure}. Originally believed to be monoclinic \cite{Nial},  the crystal structure was later corrected by single crystal X-ray diffraction and found to be best described by the space group $R\bar{3}m$ \cite{MAL76}. The Fe ions occupy the $18h$ crystallographic site (0.4953, 0.5047, 0.1131), and form bilayers of offset kagome networks. These kagome layers are in turn made up of 2 sizes of equilateral triangles, with Fe\---Fe distances of 2.732\,\AA\ and 2.582\,\AA; this is shown by the differently coloured triangles in the figure. The Fe\---Fe distance forming the bilayer is 2.584\,\AA.  The Sn ions occupy two distinct crystallographic sites, Sn1 (0.0000, 0.0000, 0.1041) and Sn2 (0.0000, 0.0000, 0.3303); the first of these lie within the kagome layers, and the second lie between the kagome bilayers. The refined values of the lattice parameters  ($a = b = 5.3147, c = 19.7025$\,\AA \ with respect to the tripled hexagonal unit cell) of the sample used in these studies are in good agreement with those of previous studies \cite{MAL76}.

5\,g of bulk \fesn\ powder were prepared by grinding stoichiometric amounts of Fe and Sn powders (purities) \cite{ICH68} in a glove box. The mixed powder was pelletised and sealed into a silica ampoule that had been put under vacuum ($10^{-5}$ mbar) and flushed out with argon three times, and then finally backfilled with argon to 3.5\,mbar in order to reduce Sn evaporation. The pelletized sample was heated to 1073\,K in a muffle furnace at 1\,K/min. At first the progress of the reaction was checked by x-ray powder diffraction (Bruker D4 Endeavor with Cu K-$\alpha$ radiation, equipped with a graphite secondary monochromator to eliminate the Fe fluorescence) every few days, and the pellets were reground, repelletized and sealed into an ampoule each time. However, it was found that the reaction is complete after 1 week and that no regrinding step is required. Each time an ampoule was removed from the furnace it was quenched by submersion into cold water, as \fesn\ is only stable between 873\,K and 1088\,K \cite{TRU70}. X-ray powder diffraction (D4 Endeavor) showed that the sample was $\sim95\%$\ Fe$_3$Sn$_2$ phase; the remainder consisted of FeSn$_2$ and FeSn phases.

Magnetic measurements were performed on \fesn\ powder using a Quantum Design MPMS-7 dc-SQUID magnetometer, and the oven insert was used for measurements above 300\,K. The dc-susceptibility was measured between 5\,K and 700\,K in fields of 100\,Oe and 1000\,Oe, and the field dependence was measured up to 10,000\,Oe at temperatures from 2\,K to 300\,K. The sample was held within a piece of aluminium foil which was attached to the end of the sample rod with copper wire for all of the studies \cite{SES07}. A straw sleeve was used at low temperature to prevent sample movement in the cryogenic gas flow.

\begin{figure}[htb]
\begin{center}
\includegraphics[scale=0.8]{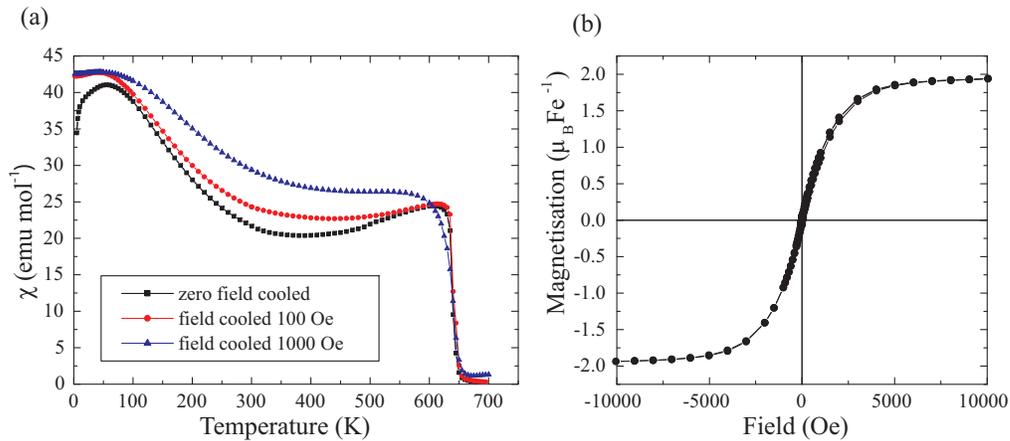}\hspace{0pc}%
\end{center}
\caption{(a) Magnetic susceptibility of \fesn\ in applied fields of 100\,Oe and 1000\,Oe, between 5\,K and 700\,K. The zero field cooled curve was measured in a field of 100\,Oe. (b) A hysteresis loop showing the field-dependence of the magnetisation of \fesn\ in fields up to 10 000\,Oe at 150\,K. }
\label{fig_susceptibility}
\end{figure}

The magnetic susceptibility ($\chi_{dc}$) of \fesn\ between 5\,K and 700\,K, in fields of 100 and 1 000\,Oe, is shown in figure~\ref{fig_susceptibility}(a). The transition into the ferromagnetic state determined from the maximum in $d\chi /dT$ {\it vs.} $T$ is  $T\mathrm{_C} \simeq 640$\,K, in fair agreement with the approximate values of 612\,K \cite{TRU70} and 657\,K \cite{CAE78} derived from M\"ossbauer data by previous workers. This transition is believed to be to a state in which the spins lie along the $c$-axis. On cooling, the 1 000\,Oe data shows that this  ferromagnetic response saturates until at  $\sim520$\,K another component causes the susceptibility to increase. We suggest that it is at this temperature that the spins begin their rotation towards the $ab$ plane. This transition is continuous until at $\sim60$\,K the susceptibility decreases. The suppression of this drop by field cooling is characteristic of a spin glass component at temperatures far below the main ferromagnetic transition, and provides further evidence of underlying magnetic frustration in this itinerant magnet. The separation of the zero-field fooled and field cooled data allow the spin glass freezing  temperature to be estimated as $T\mathrm{_f}$ $\approx$ 80\,K. It is possible that this transition is actually the onset of a second ferromagnetic component, however this seems rather unlikely as there is only one crystallographic magnetic iron site in the \fesn\ crystal structure. Field-dependent studies, shown for 150\,K in figure~\ref{fig_susceptibility}(b), indicate that there is very little coercivity and that the magnetisation  saturates in fields close to 10 000\,Oe, reaching a maximum value of $\mathrm{\sim1.9\,\mu_{B} Fe^{-1}}$. This value changes little in the range $5- 300$\,K, and is significantly less than that expected for localised Fe moments,  indicating that the Fe valence electrons are shared between localised and itinerant environments. 

\vspace{-5pc}

\begin{figure}[htb]
\begin{center}
\includegraphics[scale=0.775]{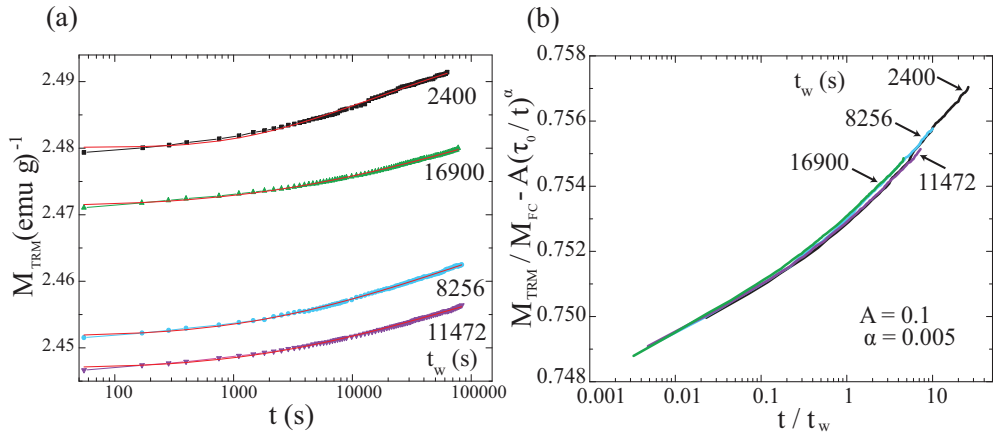}\hspace{0pc}%
\end{center}
\caption{(a) Relaxation of the thermoremanent magnetization at 25\,K, for different waiting times, $t\mathrm{_w}$. The solid lines are the best fits of the data to the equation, $M\mathrm{_{TRM}} = M\mathrm{_1} + M\mathrm{_0}\,\mathrm{exp}[-(t/\tau)^{1-n}]$. (The relative positions of the curves along the $y$ axis should not be taken as meaningful as their separations are within the error of the field produced by the SQUID.) (b) The aging part of the TRM, normalised to the field cooled value of the magnetisation as a function of $t/t\mathrm{_w}$. The stationary part, $A(\mathrm{\tau_0}/t)^{\alpha}$, where $\tau\mathrm{_0}$ is a microscopic time, has been subtracted from the data. The scaling constants $A$ and  $\alpha$ agree well with those for the AgMn spin glass \cite{VIN97}.}
\label{TRM}
\end{figure}

In order to investigate whether the low temperature transition involves a second ferromagnetic component or to a re-entrant spin glass phase, we measured the thermoremanent magnetisation (TRM) of  \fesn\  as a function of time, $t$. The sample was cooled from 300\,K to 25\,K (below $T\mathrm{_f}$) in zero field, then after a wait time, $t\mathrm{_w}$ (2400 to 16900\,s), a field of 50\,Oe was applied, and the relaxation of the TRM was measured as a function of $t$. The recorded curves, shown in figure~\ref{TRM} (a), are all well fitted by the usual function used to describe the relaxation of spin glasses: a superposition of a stretched exponential and a constant term, $M\mathrm{_{TRM}} = M\mathrm{_1} + M\mathrm{_0}\,\mathrm{exp}[-(t/\tau)^{1-n}]$ \cite{TRMref}, where $M\mathrm{_1}$ is the constant term, $M\mathrm{_0}$ is the initial TRM, $\tau$ is the characteristic time constant and $n$ is the exponent. The relaxation of the TRM in ferromagnets, on the other hand, is usually best fitted by a power law of the form $M\mathrm{_{TRM}} = M\mathrm{_1} + M\mathrm{_0}t^{-\gamma}$ \cite{TRMref}, where $M\mathrm{_1}$ is a constant, $M\mathrm{_0}$ is the initial TRM and $\gamma$ is the power law exponent. This suggests that the low temperature transition is of a spin glass nature.

The relaxation of the TRM shows a clear dependence on $t\mathrm{_w}$, which is typical in a non-equilibrium, spin-glass phase: the longer the $t\mathrm{_w}$, the slower the relaxation of the TRM \cite{H30_Fe_2,VIN97,PAR08}. Conversely, the relaxation in a ferromagnetic phase is expected to show negligible dependence on $t\mathrm{_w}$ \cite{TRMref}. If a stationary (equilibrium) part is subtracted from our TRM curves and they are plotted against $t/t\mathrm{_w}$ an almost full aging scaling is observed (figure~\ref{TRM} (b)). This further indicates that the low temperature phase transition involves a spin glass component, rather than a ferromagnetic one.

\begin{figure}[htb]
\begin{center}
\includegraphics[scale=0.45]{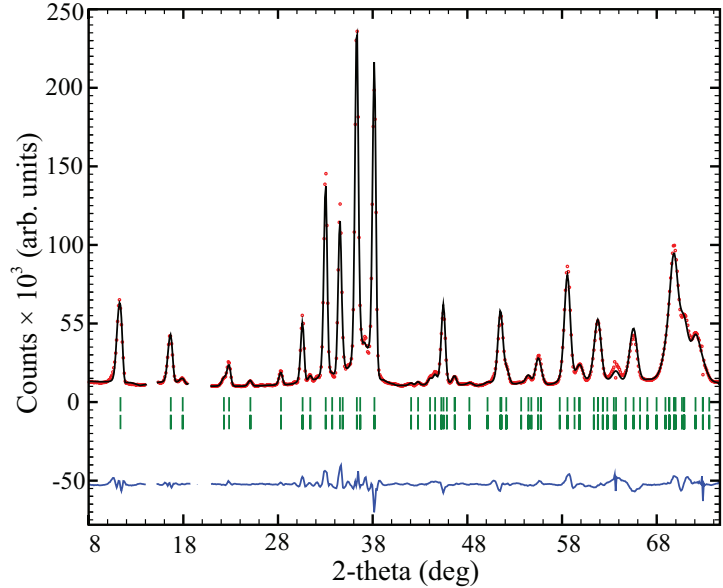}\hspace{0pc}%
\end{center}
\caption{Fit to the powder neutron diffraction pattern of \fesn. The upper tick marks indicate the positions of  peaks predicted from the nuclear phase, the lower indicate those in the magnetic phase. The circles correspond to the observed scattering, the line shows the calculated diffraction pattern and the difference is given below. Significant contamination from the cryomagnet leads to the increase in background between 30 and 38\,$^\circ$; regions where scattering from the sample environment are distinct from that of \fesn\   were excluded from the refinement. The data were collected at 300\,K using the D20 diffractometer with neutrons of wavelength 1.3\,\AA. The final goodness of fit parameters were $\chi^{2} = 113.9$\ and\ $R_{wp} = 9.87$ with 51 refined parameters.}
\label{neutron_main}
\end{figure}

In order to determine how the competing energy scales within \fesn\ are manifested in the ordering of the atomic moments, powder diffraction data were collected with neutrons of wavelength 1.3\,\AA\ using the high flux diffractometer D20 at the ILL. Approximately 2\,g of sample was held in a 10\,mm diameter vanadium can with temperature being controlled using a cryomagnet. Data were taken at 300, 150, and 6\,K (all below the Curie temperature, T$_C$ = 640\,K) in zero magnetic field. The basis vectors that describe the different symmetry types of magnetic structure were calculated using the technique of representational analysis embodied in the program SARA{\it h} \cite{SARAH}. Analysis of the crystal and magnetic structure was carried out using data over the angular range 8$^\circ \le 2\theta \le 75^{\circ}$, using Fullprof \cite{FULL} together with SARA{\it h}-Refine. 

Representational analysis indicates that the magnetic representation for the Fe crystallographic site ($18h$) is decomposed into the irreducible representations (IRs) of the little group of the propagation vector $G_k=R\bar{3}m$ according to $\Gamma_{Mag}=1\Gamma_{1}^{(1)}+2\Gamma_{2}^{(1)}+2\Gamma_{3}^{(1)}+1\Gamma_{4}^{(1)}+3\Gamma_{5}^{(2)}+3\Gamma_{6}^{(2)}$, where the subscript numbering follows that given in the works of Kovalev \cite{KovalevII} and the superscript indicates the order of the IRs. Inspection of their associated basis vectors (BVs) reveals that there are two ferromagnetic IRs with uncompensated components along the $c$-axis and in the $ab$ plane, respectively: $\Gamma_3$ and $\Gamma_5$.  Further, $\Gamma_3$ corresponds to an umbrella structure in which an ordered antiferromagnetic  component is also allowed in the $ab$ plane such that the moments are restricted to the local $ac$ mirror planes of the individual kagome triangles perpendicular to the kagome plane, a structure similar to that found in the Fe-jarosites \cite{jarosites}. $\Gamma_5$ spans 6 basis vectors (BVs) and as such corresponds to a complex magnetic structure type made up of components that are both ferromagnetic (in the $ab$ plane) and antiferromagnetic (in the $ab$ plane and $||c$).  $\Gamma_5$ also has the notable quality that it allows the moments on the different Fe-sites to be of unequal sizes. These calculations indicate a possible richness in the orderings of the Fe-moments that can occur in \fesn \ at the atomic level: a transition from a state with all the moments along the $c$-axis to one with the moments in the $ab$ plane  does not require the moments to be either collinear or equal in magnitude. As there is no symmetry requirement for the moments to be collinear and equal, it follows that the key experimental challenge is to determine the degree of non-collinearity and the variation in the moment sizes, as both of these, and the fluctuations associated with them, could lead to anomalous electron transport effects.

\begin{table}
\caption{\label{momenttable} Refined values of the moments, their angles from the principle crystallographic axes and the goodness-of-fit parameters for models of collinear and non-collinear ordering in \fesn\ as a function of temperature. All deviations correspond to standard errors, determined from the errors in the refined parameters, except for those of the angles in the non-collinear model, which correspond to the spread of angles of the individual moments.}
\lineup
\begin{tabular}{ccccccccc}
\br 
 & \multicolumn{4}{c}{Collinear} & \multicolumn{4}{c}{Non-collinear ($\Gamma_{3}\oplus\Gamma_{5}$)  }\tabularnewline \mr
T(K) & $\theta(^\circ)$ & $\phi(^\circ)$ & $\mu_{B}$(Fe)&$R_{wp}$ & $\theta(^\circ)$ & $\phi(^\circ)$ & $\mu_{B}\textrm{(Fe)} $ &$R_{wp}$ \tabularnewline \mr 
300 & 0$\pm0$ & 20.6$\pm3.6$ & 2.27$\pm0.13$ & 9.85 & 0$\pm39.9$ & 21.6$\pm\08.7$ & 2.19$\pm0.15$ & 9.87 \tabularnewline 
150 & 0$\pm0$ & 17.5$\pm8.1$ & 1.82$\pm0.18$ & 9.71 & 0$\pm83.1$ & 32.7$\pm10.9$ & 1.63$\pm0.23$ & 9.51 \tabularnewline 
6     & 0$\pm0$ & 65.9$\pm7.9$ & 1.90$\pm0.17$ & 9.79 & 0$\pm31.6$ & 68.6$\pm\06.7$ & 1.95$\pm0.70$ & 9.63 \tabularnewline \br
\end{tabular}%
\end{table}

\begin{figure}[htb]
\begin{center}
\includegraphics[scale=0.32]{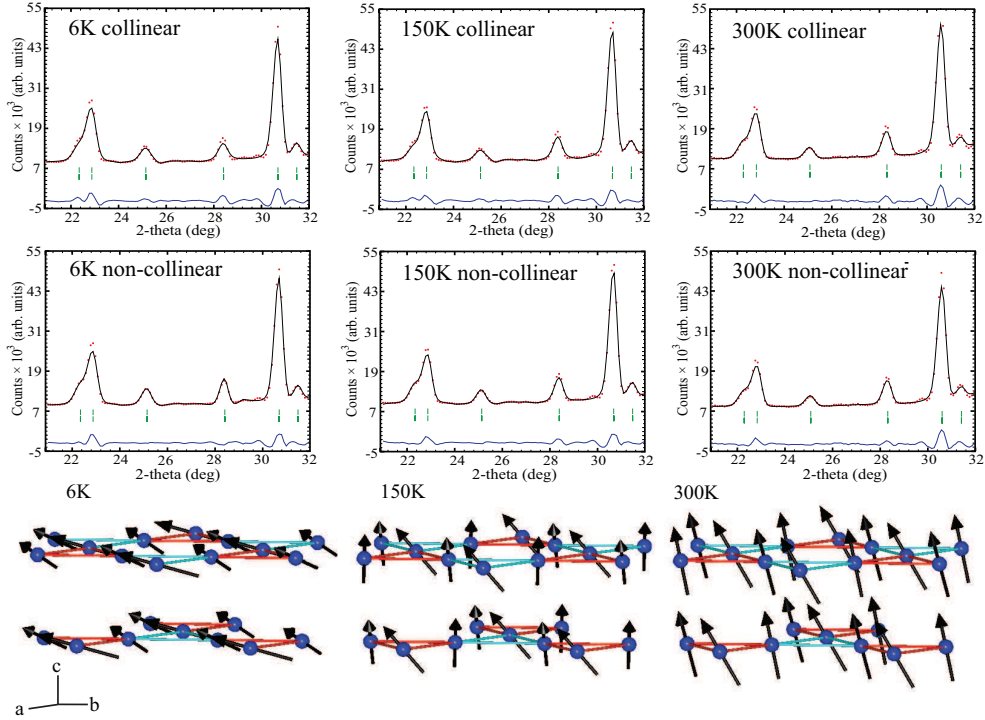}\hspace{0pc}%
\end{center}
\caption{The top panel shows the goodness of fit of the collinear magnetic structure model to the data in the region 21$^\circ \le 2\theta \le 32^{\circ}$, at 6\,K, 150\,K and 300\,K (left to right). The middle panel shows the goodness of fit of the non-collinear model in the same region at the three temperatures. The bottom panel shows the refined non-collinear magnetic structures which correspond to the plots in the middle panel.}
\label{neutron_panel}
\end{figure}

Refinement of the powder neutron diffraction data was carried out using 2 models: a simple collinear model in which the angle away from the $c$-axis of a set of identical moments was refined, and a non-collinear model in which the weighting coefficients of the  different BVs calculated by representational theory were refined. Both models indicate that at 300\,K (figure~\ref{neutron_main}) the moments lie largely along the $c$-axis (\ref{momenttable}). There is essentially no difference between the quality of the fits for the collinear and non-collinear ($\Gamma_3 \oplus \Gamma_5$) models, and in both cases the deviation of the average moment direction from the $c$-axis is approximately  $\phi\sim\pm21^\circ$. On cooling, both models show the moments to be flopping into the $ab$ plane, with only a small discrepancy appearing in their ability to fit the experimental diffraction data: the  magnetic scattering at $\simeq 25.1^{\circ}$ and $\simeq 28.3^{\circ}$ is better fitted by the non-collinear structures of $\Gamma_3 \oplus \Gamma_5$ at both 6 and 150\,K (figure~\ref{neutron_panel}).  The observation of ordered Fe-moments of $\sim2$\,$\mu_B$ is in good agreement with prior M\"ossbauer \cite{CAE78} and powder diffraction studies \cite{MAL78}. No changes in the average refined magnitude of the moments were observable in this experiment, indicating that this is not the main drive for the spin reorientation transition. Further, the similarities between the magnetic diffraction patterns and  refined models at 6 and 150\,K indicate that the spin structures in the intermediate and spin glass phases are closely related. Unfortunately, the large background from the sample environment prevents any comment from being made about the strength of the  diffuse scattering associated with the disordered spin glass component, and how it changes upon cooling.

The equivalence in the quality of the fits from the collinear and non-collinear models indicate that unpolarised powder neutron diffraction does not have the sensitivity required  to unambiguously pin down the degree of canting together with the variation in the sizes of the magnetic moments.  Previous attempts to improve the quality of the fit of the collinear model through modification of the Fe-form factor are not well justified \cite{MAL78} and lead to a unsatifying model. Rather, we argue that the non-collinear model is to be preferred as it allows resolution of difficulties in the interpretation of early $^{57}$Fe and $^{119}$Sn M\"ossbauer  data \cite{CAE78,CAE79}, and a consistent picture of the temperature-dependent spin transition of \fesn\ to be constructed. 

In the early studies the authors concluded that the magnetic structure features 2 components. The first has population $\alpha$ and  is a collinear ferromagnetic component where the moments lie almost parallel with the {\it c}-axis above 250\,K, gradually rotate towards the $ab$ plane on cooling below 250\,K, and lie in the $ab$ plane at low temperature, remaining collinear throughout the rotation. A second contribution was required to model the rotation of the moments on warming; it involves moments in the $ab$ plane and has a population $(1-\alpha)$ that decreases slowly on warming. Our model of non-collinear ferromagnetism allows an alternative interpretation of these data: the spins continuously rotate from the $ab$ plane to the $c$ direction up to $\sim$ 520\,K, and feature a non-collinear component that is temperature dependent. The slow rotation then indicates that the  near balance of the energy scales responsible is temperature insensitive. The spin glass transition at low temperature ($T\mathrm{_f}\simeq 80$\,K) is then to a phase with the moments largely within the $ab$ plane, a situation reminiscent of the anisotropy-induced spin glass state of the kagome antiferromagnet (H$_3$O)Fe$_3$(SO$_4$)$_2$(OH)$_6$ \cite{H3O_anisotropy,H3O_Fe}.

In conclusion, we show that \fesn\ is a rare example of a frustrated itinerant magnet. Three transitions are observed upon cooling: the first at $T_C = 640$\,K is from the paramagnetic phase to a collinear ferromagnetic phase with the moments collinear with the $c$-axis. On cooling from $\sim520$\,K to $\sim75$\,K the moments rotate from the $c$-axis into the $ab$ kagome plane. During this transition symmetry restrictions that require the moments to be collinear and of equal size are relaxed, allowing a non-trivial ferromagnetic structure to develop. The energy scales responsible for this spin structure are at present unclear, though the Dzyaloshinsky-Moriya interaction \cite{D-M}, which is allowed on the kagome lattice, is an obvious candidate. \cite{Elhajal}.

On further cooling below $\sim 75$\,K the competition between magnetic interactions leads to a transition to a re-entrant spin glass phase. The origins of this spin glass phase are unclear as such behaviour is more commonly observed in highly disordered ferromagnets, {\it e.g.} Fe$_{0.7}$Al$_{0.3}$ \cite{Fe0.7Al0.3}, whereas \fesn\ is not a disordered system. Further work is also required to understand the role that the high degree of frustration and the 2-dimensional fluctuations expected from the underlying kagome lattice play in the magnetism of this material.

Our studies on \fesn\  also indicate its potential for use in spintronics for both spin injection and applications based on the AHE. The latter may be enchanced above the values expected for a conventional ferromagnet by a coupling  between the conduction electrons and the non-trivial ferromagnetic spin structure. Such a coupling is likely as  a large Hall resistivity has recently been observed at room temperature in the granular alloy films with the composition Fe$_{68}$Sn$_{32}$  of around 60 times greater than the Hall resistivity of pure Fe, though details of the underlying magnetic structure are not currently known \cite{GAO03}.

\end{document}